\documentstyle[aps,prl,floats,twocolumn,amsfonts,amssymb]{revtex}
\input epsf
\draft

\begin{document}

\renewcommand{\bottomfraction}{0.95}
\renewcommand{\topfraction}{0.95}
\renewcommand{\textfraction}{0.1}
\renewcommand{\floatpagefraction}{0.7}
\renewcommand{\thesection}{\arabic{section}}

\addtolength{\topmargin}{10pt}


\newcommand{\eq}{\begin{equation}}
\newcommand{\en}{\end{equation}}
\newcommand{\eqa}{\begin{eqnarray}}
\newcommand{\ena}{\end{eqnarray}}
\newcommand{\eqan}{\begin{eqnarray*}}
\newcommand{\enan}{\end{eqnarray*}}
\newcommand{\spz}{\hspace{0.7cm}}
\newcommand{\lbl}{\label}


\def\Bbb{\mathbb}


\newcommand{\Dslash}{{\slash{\kern -0.5em}\partial}}
\newcommand{\Aslash}{{\slash{\kern -0.5em}A}}

\def\sqr#1#2{{\vcenter{\hrule height.#2pt
     \hbox{\vrule width.#2pt height#1pt \kern#1pt
        \vrule width.#2pt}
     \hrule height.#2pt}}}
\def\smallsquare{\mathchoice\sqr34\sqr34\sqr{2.1}3\sqr{1.5}3}
\def\square{\mathchoice\sqr68\sqr68\sqr{4.2}6\sqr{3.0}6}
 
\def\thinspace{\kern .16667em}
\def\punto{\thinspace .\thinspace}
 
\def\xp{x_{{\kern -.2em}_\perp}}
\def\subp{_{{\kern -.2em}_\perp}}
\def\kperp{k\subp}

\def\derpp#1#2{{\partial #1\over\partial #2}}
\def\derp#1{{\partial~\over\partial #1}}

\def\zbar{\overline{z}}
\def\wbar{\overline{w}}
\def\ez{{{\bf e}}_z}
\def\ezbar{{{\bf e}}_{\zbar}}
\def\vF{ v_{_{\rm F}} }
\def\EF{ E_{_{\rm F}} }
\twocolumn[\hsize\textwidth\columnwidth\hsize\csname@twocolumnfalse%
\endcsname

\title{Stochastic Heterostructures in B/N-Doped Carbon Nanotubes}

\author{Paul~E.~Lammert$^1$\thanks{lammert@phys.psu.edu},
Vincent~H.~Crespi$^1$, and Angel~Rubio$^2$}
\address{$^1$Dept. of Physics, 
The Pennsylvania State University \\
104 Davey Lab, University Park, PA 16802-6300}
\address{$^2$
Departamento de F{\'\i}sica de Materiales, Facultad de
Ciencias Qu{\'\i}micas, Universidad del Pais Vasco/Euskal Herriko
Unibertsitatea, Apdo. $1072, 20018$ San Sebasti\'an/Donostia, Basque
Country, Spain}

\date{\today}
\maketitle

\begin{abstract}

\widetext
Carbon nanotubes are one-dimensional and very narrow.
These obvious facts imply that under doping
with boron and nitrogen, microscopic doping inhomogeneity 
is much more important than for bulk semiconductors.  We consider 
the possibility of exploiting such fluctuations to create 
interesting devices.  Using self-consistent tight-binding 
(SCTB), we study heavily doped highly compensated 
nanotubes, revealing the spontaneous formation of structures 
resembling chains of random quantum dots, or nano-scale
diode-like elements in series.  We also consider truly isolated 
impurities, revealing simple scaling properties of bound state
sizes and energies.

\end{abstract}
\vskip0.3cm

]
\narrowtext

The intrinsic transport properties of carbon nanotubes are 
promising for nanoelectronic applications, but useful 
intra-tube devices usually require electronic heterogeneity.
To date, experimenters have either imposed the requisite 
heterogeneity through electron-beam lithography on a much 
larger length scale ({\it e.g.,} gate contacts, masked potassium 
doping\cite{Kdoped}), or exploited preexisting 
inhomogeneities such as on-tube defects\cite{M/SC},
chance placement of external nano-particles, or tube-tube 
crossings\cite{crosses}.
One potential route to stable, small-scale heterogeneity 
of electronic properties is substitutional doping by boron and 
nitrogen.  Such doping has already been achieved by a variety
of techniques, including
arc discharge\cite{synth_arc},
pyrolysis\cite{synth_pyro}, chemical vapor
deposition\cite{synth_cvd}, and substitution
reactions\cite{synth_sub_1}. 
Spatial variation in dopant density is
much more important in 1-D nanotubes than in bulk semiconductors.
Fluctuations are large and electrons cannot bypass anomalous regions.
Normally, such fluctuations would be considered a nuisance
or worse.  
We ask: Can they be exploited?
After all, a heavily doped compensated nanotube is 
a chain of p/n junctions (which might be called ``diodium'').
We demonstrate the potential of the idea 
by studying the electronic structure of heavily doped 
($\sim 1$ atomic \%), highly compensated nanotubes. 

This paper takes an atomic-level view of doping and junction-like 
structures.  
Previously, large-scale inhomogeneous structures, 
{\it i.e.,} Schottky barriers and p-n junctions, were investigated in 
continuum\cite{leonard-tersoff-novel,odintsov,%
esfarjani-farajian} and self-consistent virtual 
crystal\cite{leonard-tersoff-negative} approximations, with
an emphasis on the meso-scale electrostatic peculiarities
of one-dimensional systems\cite{leonard-tersoff-novel}.
Such studies provide valuable insights, but cannot access
properties arising from atomic-scale structure.
{\it Ab initio\/} techniques, on the other hand, have been 
applied to a variety of microscopically ordered, homogeneous, 
nanotube materials: 
BN, BC$_3$, and BC$_2$N\cite{rubio-et-al};
BC$_{40}$ and NC$_{40}$\cite{yi-bernholc},
C$_{17}$N\cite{czerw}, and abrupt C/BN junctions\cite{blase-recursion}.
There is no previous theoretical study of genuinely randomly 
({\it i.e.,} realistically) doped and compensated nanotubes.

In a $d$-dimensional doped semiconductor, 
the {typical} fluctuation in total dopant charge in a region of linear 
size $L$ is proportional to $L^{d/2}$ (assuming Gaussian statistics), 
leading to fluctuations in the unscreened Coulomb potential 
proportional to $L^{(d-2)/2}$.  Only for $d=1$ is the exponent 
$(d-2)/2$ negative.  This implies that the dominant fluctuations 
are then on the shortest length scale over which the system is 
genuinely one dimensional, {\it i.e.,} the circumference.
(If the doping is so low that the length per dopant 
[$\lambda$] exceeds the circumference, then $\lambda$ is the 
dominant scale.)
This contrasts with bulk materials, for which the most
important fluctuations can be on very long length scales 
and are controlled by screening, especially for light doping 
and high compensation\cite{shklovskii-efros}.
Screening is ineffective at short distances in nanotubes;
taking screening into account merely strengthens the conclusion.
We can estimate the potential
fluctuation ($\Delta V$) over an axial distance $L = 2\pi R$
by computing the average potential produced by
a Gaussian (``$\sqrt N$'') charge fluctuation.
Denoting the proportion of atoms which are dopants by $c$,
the dielectric constant by ${\tilde{\epsilon}}$\cite{epsilon},
and using the graphene lattice constant $2.45$\AA, we find
\hbox{$\Delta V \approx$ 9 eV $c^{1/2}/{\tilde{\epsilon}}$}.
This is consistent with the numerical results presented below.
Note that $\Delta V$ is independent of $R$ for fixed $c$.
Fluctuations in the azimuthal locations of dopants are 
not important: all azimuthal Fourier components of 
the doping have the same variance, but the potential 
arising from nonuniform components 
($m\not= 0$, see Eq. \ref{Vm}) falls off 
rapidly and cannot be resolved by intra-band processes.
(Numerical results confirm that they have little effect.)

\begin{figure}
\epsfxsize 92 mm
\epsfbox{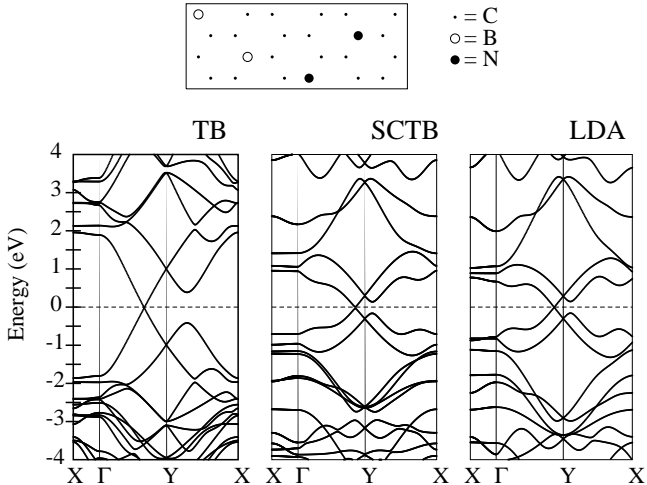}
\caption{Test of SCTB against LDA on a regular planar BC$_{10}$N structure.
Top: the unit cell.  Bottom: Electronic energy bands near
the Fermi level (dashed line) calculated by simple TB, SCTB, and LDA-DFT.
}
\label{comparison}
\end{figure}

{\it Ab initio\/} methods are prohibitively expensive 
for large systems, and ordinary tight binding is unable 
to treat charge transfer over distances greater than 
about a bond length, so that we have chosen to
use a self-consistent tight-binding (SCTB) technique.
The nonorthogonal\cite{nonorth} tight-binding basis contains four orbitals 
per atom with Hamiltonian and overlap parametrization 
determined directly from LDA\cite{porezag}.
This parametrization was shown to give results in good 
agreement with LDA over a wide range of structures with
short-range charge transfer.
We impose self-consistency of long-wavelength components of 
the Hartree Coulomb potential {\it via\/} an 
iterative procedure\cite{unpub}.
Figure \ref{comparison} compares results of simple TB,
self-consistent TB, and LDA for a strongly charge modulated
planar structure.
SCTB is remarkably effective at correcting the 
deficiencies of simple TB, reproducing the main band-narrowing
and shift of Fermi energy level crossing exhibited by the LDA
calculation, and is therefore expected to describe the physics
near $\EF$ well.
The data presented here are for mechanically unrelaxed systems.  
Tests show that relaxation introduces only minor corrections 
to the electronic structure.

\begin{figure}
\epsfxsize 92 mm
\epsfbox{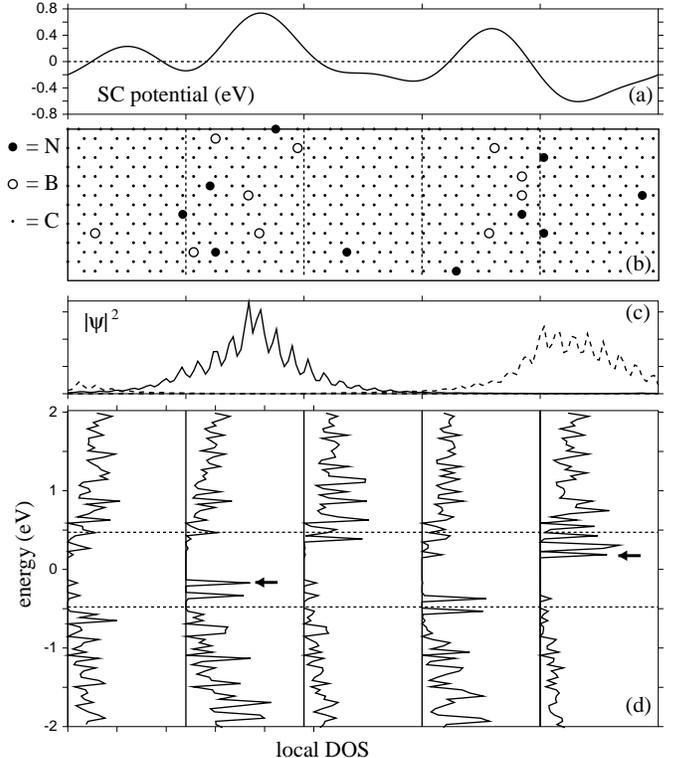}
\caption{SCTB results for sample A,
an (8,0) nanotube randomly doped with 10 each nitrogen and boron
atoms per supercell of length 7.7 nm (overall doping level 3.5\%).
The horizontal direction in all figures corresponds to the axial direction.
(a) Azimuthally averaged self-consistent potential.
(b) The doping pattern in the unit cell cut open and laid flat.
(c) Axial probability density profiles of deep donor
and acceptor states (marked by arrows in the DOS plots).
(d) Total densities of states for the corresponding slices 
indicated by dashed lines in panel (b).  Dashed lines show the
gap edges of an undoped (8,0) nanotube (centered at zero).
}
\label{sample A}
\end{figure}
\begin{figure}
\epsfxsize 92 mm
\epsfbox{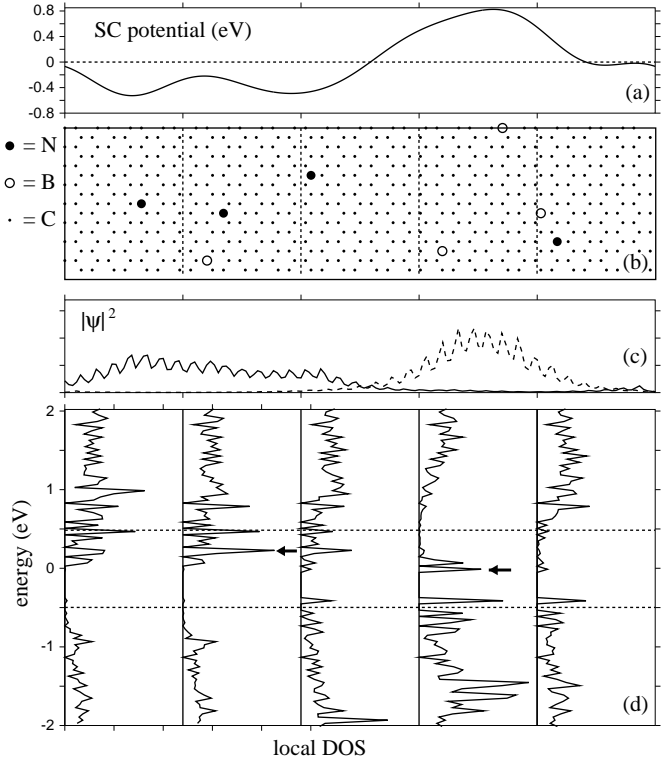}
\caption{Sample B.  SCTB results for an (8,0) nanotube with a
better segregated and lower density arrangement than sample
A, containing 4 each nitrogen and boron atoms per 7.7 nm unit cell.
Presentation is as in Fig. \ref{sample A}.
}
\label{sample B}
\end{figure}

A length of nanotube heavily doped with both boron and 
nitrogen might have a substitution pattern like that shown in
the second panel of Figure \ref{sample A}.
This shows a section of (8,0) nanotube (cut open and
rolled flat with the axial direction horizontal)
containing a number of boron and nitrogen atoms substituted on 
randomly chosen sites.
The section is treated as the unit cell of a superlattice.  
A more lightly doped sample is shown in figure \ref{sample B}.
Fluctuations in the doping act as {\em collective impurities}, 
creating localized impurity states in the energy 
gap of the undoped nanotube.
This localization is illustrated by the plots of the axial 
probability densities (azimuthal average of $|\psi|^2$) of 
deep collective donor and acceptor states.
These states arise from a local imbalance in the dopant 
charge of only one or two unit charges.
Regions with excess nitrogen atoms suppress the 
local density of states near the bottom of the gap 
({\it vice versa\/} for boron-rich regions).  
This remnant of band bending would matter little 
in a bulk material, but in a one-dimensional structure can 
effectively block charge transport if the Fermi level 
lies in such a local pseudogap.  

These randomly doped nanotubes resemble disordered chains of 
tiny quantum wells or dots, and could show a variety of 
behaviors characteristic of such structures, {\it e.g.}, 
negative differential resistance.  
Level spacing in the wells is $\Delta \sim 10^{-1} E_{\rm gap}$.  
Band widths for the well states, $\Gamma \sim 1$ meV 
for the samples studied, measure interwell tunneling matrix 
elements.
The charging energy scale, $E_C = e^2/({\tilde{\epsilon}}2\pi R) = 
(0.64/{\tilde{\epsilon}})E_{\rm gap}$, scales with the gap.
This is a significant energy, so that one must remember the
local DOS plots in the Figures are specific for neutrality in
the depicted unit cells.
A length of compensated doped semiconducting tube,
particularly under imperfect dopant mixing, forms a 
dense sequence of alternately directed nanoscale diode-like
elements. 
Reverse-biased structures should dominate transport under
a voltage; field doping the tube would wash out weaker junctions,
isolating regions of most pronounced nonlinear behavior.

Leonard and Tersoff\cite{leonard-tersoff-novel} 
provided a formula showing that the depletion layer 
of a nanotube p/n junction is very long at low 
doping, but shrinks exponentially with doping.  
Unfortunately, the formula is not applicable to the high doping 
($ \sim 1$\%) regime relevant to devices with nanometer-scale 
structure.  The depletion width in sample B is close to being 
saturated and is very narrow ($\sim 1$ nm).  

\begin{figure}
\epsfxsize 92 mm
\epsfbox{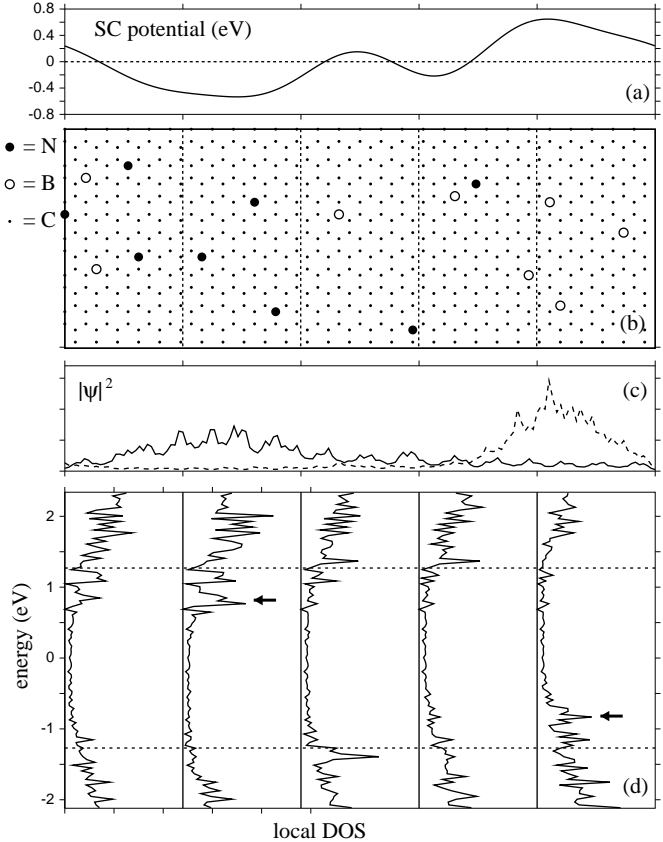}
\caption{Sample C.  SCTB results for a doped (6,6) nanotube 
(6.9 nm unit cell) with 1.2\% each boron 
and nitrogen.  The dashed lines, centered at zero, indicate the edges
of the $m = \pm 1$ subbands of the undoped tube, at which
the DOS jumps suddenly from a low constant value.
}
\label{sample C}
\end{figure}

Remarkably, many features of this picture also hold for
doped metallic nanotubes, as shown 
in Figure \ref{sample C}. The Fermi level of a pristine 
metallic tube lies at the center of a subband which 
contributes a small and constant density of states.
Upon doping, this subband forms a relatively inert background 
over which appears a response very similar to that of 
a semiconducting tube.
One observes band bending and highly concentrated 
(though not technically localized) states in the would-be gap.
The roughening of the local DOS near the Fermi level
seen in the figure is partly an artifact of periodicity
and k-point sampling, but it also suggests that 
backscattering may be stronger than in an 
extrapolation from the limit of vanishing doping\cite{choi}.

We lastly consider states bound to isolated impurities,
using both SCTB and an effective mass approximation.
These studies reveal new diameter-dependent scaling relations\cite{reliable}.
A surface charge density 
$\rho_m =  ({Q_m}/{2\pi}) \delta(z) e^{im\phi}$ 
on a cylinder of radius $R$ ($\phi$ is the azimuthal angle,
$z$, axial position)
produces a potential at $(z,\phi=0)$ which falls off 
asymptotically 
($z \to \infty$) as 
\begin{equation}
V_m(z) \sim 
\frac{(2|m|)!}{2^{3|m|}(|m|!)^2}
\frac{Q_m/{\tilde{\epsilon}}}{\sqrt{z^2 + 2R^2} } 
\left[ \frac{z^2}{2R^2} + 1 \right]^{-|m|}, \,\,
\label{Vm}
\end{equation}
This familiar result of multipole expansion shows that only the 
$m=0$ component of charge has an influence of any appreciable range.
We retain only the $m=0$ component of the impurity Coulomb field, 
partly due to this estimate, but mostly because 
$m\not= 0$ contributes only if there is band mixing.  
The $m=0$ component of the potential can be written explicitly
as an integral, but it differs little from the potential
of an identical charge located on the tube axis, so we
use the latter for simplicity, taking the charge to be
$Z |e|$.
Ignoring the trigonal distortion of the nanotube electronic structure,
the band-slicing picture gives an effective mass
$ m^* = E_{\rm gap}/(2\hbar^2 \vF^2)$,
with which we obtain the one-dimensional wave equation 
\begin{equation}
\frac{d^2\psi}{d\zeta^2} = -\left( \frac{E_{\rm gap}R}{\hbar\vF}\right)^2
 \left[ \frac{E}{E_{\rm gap}}+ 
\frac{Z e^2}{{\tilde{\epsilon}} R E_{\rm gap}} 
\frac{1}{\sqrt{1+\zeta^2}} \right] \psi,
\label{wave}
\end{equation}
where $\zeta = z/R$ is the axial coordinate scaled by the tube radius.
The bandgap of a large-gap nanotube (one whose wrapping indices do 
not differ by a multiple of three)
satisfies ${E_{\rm gap}}R = 2 \hbar \vF/3$.  Therefore, if 
$Z/{\tilde{\epsilon}}$ is held fixed, both $E/E_{\rm gap}$,
the ratio of bound state energy to the gap, and 
$\psi(z/R)$ are independent of tube radius. Surprisingly,
the bound state size and energy scale with the diameter
of the tube.

The energies and axial spreads of the bound states
are determined numerically from Eq. (\ref{wave}).  
Empirically, both scale extremely
well with $Z/{\tilde{\epsilon}}$ between 1 and 10.
We find
\begin{eqnarray}
E/E_{\rm gap} & \approx &  -2.65 (Z/{\tilde{\epsilon}})^{1.25},
\nonumber \\
\langle (z/R)^2 \rangle & \approx & 1.2 (Z/{\tilde{\epsilon}})^{-1.05}.
\label{eff-mass}
\end{eqnarray}
These results are consistent with self-consistent tight-binding.
Our SCTB calculation for an (8,0) nanotube with one 
nitrogen or boron impurity per 7.7 nm unit cell verifies
the existence of a bound state fairly well localized within 
that region ($\langle \Delta z \rangle \simeq  3.2 R$).  
Since the calculation assumes a periodicity,
the average potential in the unit cell does
not correspond to that of an isolated impurity,
so we must subtract the relevant offset.
This procedure is somewhat uncertain since $\tilde{\epsilon}(k)$ 
varies from 7 to 4.5 as $kR$ falls from  $1.0$ to $0.3$.
Assuming $k = 2\pi/$({cell length})
gives a binding energy of about $0.23 E_{\rm gap}$.
This value is duplicated by Eq. (\ref{eff-mass}) 
with $\tilde{\epsilon} = 7$, which is reasonable 
due to the importance of the short-range part of the potential.
With $\tilde{\epsilon} = 7$, the model predicts a spread
$\langle \Delta z \rangle \simeq 3R$, very close to the SCTB 
result.
 
In summary, the pointlike character of dopant charges 
in B/N substitutionally doped nanotubes is important
at all doping levels.  
At high doping, fluctuations spontaneously produce a high 
density of nonlinear nanometer-scale quantum dot 
and/or junction structures, which are relatively insensitive 
to environmental perturbations such as packing into nanotube 
bundles (note that these junctions do not simply provide 
scaled-down traditional functionality).  
These results are also relevant to chemisorbed dopants.
The one-dimensional nature, which makes doping variation 
especially relevant to electronic characteristics,
also facilitates mapping and manipulation, 
{\it e.g.,} by scanning tunneling microscopy.  
Using fluctuations in this way can produce nonlinear 
device characteristics on scales far smaller than can 
be fabricated intentionally and avoids the need for 
fine doping control.

AR acknowledge financial
support from the European RTN network contract HPRN-CT-2000-00128
(COMELCAN), RTD-FET contract IST-2000-26351 (SATUNET) and JCyL (VA28/99).
PL and VC acknowledge support through National Science Foundation 
grant DMR--9876232.


\begin{thebibliography}{}
\bibitem{reliable}
It is hard to draw solid conclusions about isolated 
impurities from previous studies (such as ref. \cite{yi-bernholc})
since fairly high doping levels were used.
\end{thebibliography}
\end{document}